\documentclass{ifacconf}
\usepackage{color}
\usepackage{graphics}
\usepackage{graphicx} 
\usepackage{natbib}     
\usepackage{subfigure}
\usepackage{amsmath}
\usepackage{amssymb}
\usepackage{algorithm}
\usepackage{algorithmic}
\usepackage{enumitem}
\setitemize[1]{itemsep=5pt,partopsep=1pt,parsep=\parskip,topsep=5pt}

\DeclareMathOperator*{\argmin}{arg\,min}
\definecolor{philipp}{rgb}{0.0,0.5,0.0}

\begin{document}

\begin{frontmatter}
	
\title{Distributed Control Enforcing Group Sparsity 
in Smart Grids\thanksref{footnoteinfo}} 
\thanks[footnoteinfo]{The first two authors contributed equally. YJ, PS and KW are supported by the German Federal Ministry for Education and Research (BMBF; Grant 05M18SIA). BH and YJ are supported by ShanghaiTech University under Grant No. F-0203-14-012. KW is also indebted to the German Research Foundation (DFG-grant WO 2056/6-1).}

\author[First]{Philipp Sauerteig}
\author[First,Second]{Yuning Jiang} 
\author[Second]{Boris Houska} 
\author[First]{Karl Worthmann}

\address[First]{Institute for Mathematics, Technische Universit\"at Ilmenau, Ilmenau, Germany \\
(email: [philipp.sauerteig, karl.worthmann]@tu-ilmenau.de)}
\address[Second]{School of Information Science and Technology, ShanghaiTech University, Shanghai, China \\
(email: [jiangyn, borish]@shanghaitech.edu.cn)}

\begin{abstract} 
In modern smart grids, charging of local energy storage devices is coordinated  
on a residential level to compensate the volatile aggregated power demand on the time interval of interest. However, this results in a perpetual usage of all batteries which reduces their lifetime. 
We enforce group sparsity by using an $\ell_{p,q}$-regularization on the control to counteract this phenomenon.
This leads to a non-smooth convex optimization problem, for which we propose a tailored Alternating Direction Method of Multipliers 
algorithm. We elaborate further how to embed it in a Model Predictive Control 
framework. 
We show that the proposed scheme yields 
sparse control while achieving reasonable overall peak shaving by 
numerical simulations.
\end{abstract}


\end{frontmatter}

\section{Introduction}
The energy transition comes along with a fundamental transition of energy networks from centralized to decentralized power generation. As a result, massive storage devices have been installed in the grid to compensate the volatile generation via renewable energies and the accompanied bidirectional power flow. This paradigm shift in energy supply comes along with great optimization potential~(\cite{Lezhniuk2019,Atzeni2013,Bolognani2013}). In~(\cite{Hubert2011}) the authors discussed the importance of optimization algorithms on a residential level. The potential of coordinating local energy storages to achieve an overall goal was illustrated in~(\cite{Worthmann2015}). In~(\cite{Atzeni2013}), a game theory based approach has been proposed in order to solve a (non-)cooperative optimization problem, which arises in demand-side optimization in smart grids. The authors design distributed optimization algorithms to optimally exploit the residential generation and/or storage devices. 
The grid operator has to compensate the fluctuations of the aggregated power demand profile caused by renewable energies
~(\cite{Morstyn2018a}). Hence, one of the main goals in smart grid optimization is to coordinate the local energy storage devices at the household level in such a way that these fluctuations are mitigated.  

State-of-the-art method to tackle optimal control problems in a receding horizon fashion is Model Predictive Control (MPC), see e.g. (\cite{Worthmann2015}) for an MPC approach in smart grids. 
Efficiently solving the inherent large-scale optimization problem online is crucial for designing a practical MPC scheme. To this end, distributed optimization algorithms exploiting the network structure have been widely applied~(\cite{Boyd2011,Braun2018,Houska2016}). In practice, distributed algorithms can achieve optimality by solving local problems in parallel and only exchanging certain data with a superordinate unit. Thus, distributed optimization builds a bridge between centralized and decentralized optimization~(\cite{Worthmann2015}). A classical approach is based on dual decomposition. A class of these techniques use first-order methods to solve the corresponding dual problem~(\cite{Rantzer2009,Richter2011}). Alternatively, semi-smooth Newton methods are applied, which require a line search sub-globalization routine~(\cite{Frasch2015}). 
A centralized consensus variant of the Alternating Direction Method of Multipliers (ADMM) has been applied to solve a convex  
control problem in smart grids in~(\cite{Braun2018}). 
For an introduction to ADMM we refer to~(\cite{Boyd2011,Hong2017}). 
The approach in~(\cite{Braun2018}) and related works, however, enforces all batteries within the grid to work non-stop. This (dis)charging behavior reduces the batteries lifespan~(\cite{Ng2009,Gee2013}). We aim to counteract this effect by enforcing sparsity of the optimal control.

Sparse optimal control has already been proposed to reduce the use of batteries: In~(\cite{Salem2017}) the authors use an ultra sparse matrix rectifier for battery charging while in~(\cite{Jain2018}) an $\ell_1$ penalty is applied to enforce sparse communication of the linear quadratic regulator. A Newton-type method proposed in~(\cite{Polyak2019}) was designed for solving a general sparse optimal control problem. However, these methods do not consider the scenario of numerous batteries incorporated in a structured network as in our application. 

In this paper, we exploit group sparsity in order to extend the lifetime of the batteries. First Section~\ref{sec:problem_formulation} recaps the basic model of residential energy systems proposed in~(\cite{Worthmann2014}). Then, our main contribution is proposed, which enforces sparsity of the optimal control by introducing $\ell_{p,1}$ regularization terms. Here, $p \in \{1,2\}$ denotes the choice of the sparsity pattern. This $\ell_{p,1}$ regularization was introduced in~(\cite{Yuan2006}) in order to select grouped variables for accurate prediction in statistical learning. The $\ell_{p,1}$ regularization couples the control of each subsystem in time, which differs from the classical additive control regularization. Section~\ref{sec:distributed_MPC} presents another contribution, which elaborates how to design and implement an ADMM based optimization scheme to solve the underlying optimization problem online in a distributed manner. Our numerical open-loop results in Section~\ref{sec:numerics} show the potential of this approach to prolong the batteries' lifespans while achieving a reasonable overall performance. Moreover, we investigate the closed-loop performance in an MPC scheme numerically. Finally, We conclude the paper with Section~\ref{sec:conclusions}.

\emph{Notation:} Throughout this paper we use the notation $\mathbb{N} = \{1,2,\ldots\}$ and $\mathbb{N}_0 = \mathbb{N} \cup \{0\}$. Furthermore, we denote $[m:n] := [m,n] \cap \mathbb{N}_0$ for $m,n\in \mathbb{N}_0$ with $m \leq n$ and $\mathbf{1}_\ell = (1, 1, \ldots, 1)^\top \in \mathbb{R}^\ell$ for $\ell \in \mathbb{N}$; $\mathbf{0}_\ell$ is defined analogously. The Kronecker product of two matrices $A \in \mathbb{R}^{k \times l}$ and $B \in \mathbb{R}^{m \times n}$ is given by $A \otimes B = \left(a_{ij}B\right)_{i,j} \in \mathbb{R}^{km \times ln}$.

\section{Sparse Control Problem}
\label{sec:problem_formulation}
This section recaps the basic model of smart grids consisting of residential energy systems, which are coordinated by a grid operator in order to achieve an overall optimum as described in~(\cite{Worthmann2014,Ratnam2017}). Furthermore, we introduce a regularization in order to establish group sparsity with respect to the batteries and thus, prolong their lifespans. At the end, the resulting optimal sparse control problem is addressed as a distributed non-smooth convex optimization problem. 

\subsection{Residential energy systems}
\begin{figure}[htbp!]
	\centering
	\includegraphics[width=0.35\textwidth]{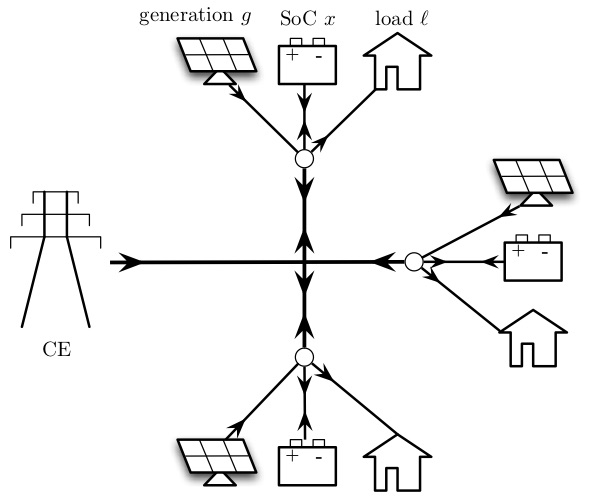}
	\caption{Network of residential energy systems.}
	\label{fig::Network}
\end{figure}

Let us consider a smart grid with $\mathcal{I}$, $\mathcal{I} \in \mathbb{N}$, residential energy systems. Each subsystem incorporates its load, power generator, and battery serving as the energy storage device. As shown in Figure~\ref{fig::Network}, the grid operator acts as a Central Entity (CE) compensating the local power demands.

The $i$-th subsystem, $i \in [1:\mathcal{I}]$, is described by 
\begin{subequations}
\label{eq::dynRES}
\begin{align}\label{eq::dynRES1}
x_i(n+1) & = \alpha_i x_i(n) + T(\beta_i u_i^+(n) + u_i^-(n))\;,\\\label{eq::dynRES2}
z_i(n) & = w_i(n) + u_i^+(n) + \gamma_i u_i^-(n)\;. 
\end{align}
\end{subequations}
State $x_i(n)$ and control inputs $u_i(n)=\begin{pmatrix}u_i^+(n) & u_i^-(n)\end{pmatrix}^\top$ denote the State of Charge (SoC) in~kWh, and the charging and discharging rate in~kW at time instant $n \in \mathbb{N}_0$, respectively. The output $z_i(n)$ denotes the power demand in~kW, and $w_i(n)$ is the net consumption (load minus generation) in~kW. The parameter $T>0$ represents the length of the time interval in~h, i.e., $T=0.5$ corresponds to $30$~min, while $\alpha_i,\beta_i,\gamma_i \in (0,1]$ model efficiencies with respect to self-discharge and energy conversion, respectively. 

The SoC and the (dis-)charging rate are subject to the constraints
\begin{subequations}\label{subeq:constraints}
	\begin{eqnarray}
	\label{eq:constraints:soc} 0 \quad \leq \quad & x_i(n) & \leq \quad C_i \;,\\
	\label{eq:constraints:discharge} \underline{u}_i \quad \leq \quad & u_i^-(n) & \leq \quad 0 \;,\\
	\label{eq:constraints:charge} 0 \quad \leq \quad & u_i^+(n) & \leq \quad \overline{u}_i\;,\\
	\label{eq:constraints:chargeANDdischarge} 0 \quad \leq \quad & \frac{u_i^-(n)}{\underline{u}_i} + \frac{u_i^+(n)}{\overline{u}_i} \quad & \leq \quad 1\;,
	\end{eqnarray}
\end{subequations}
where $C_i \geq 0$ denotes the battery capacity. When charging and discharging occurs during one time interval, condition~\eqref{eq:constraints:chargeANDdischarge} is introduced such that the box constraints on $u_i^+$ and $u_i^-$ still hold. Note that we allow $z_i$ to be negative, i.e., the subsystems are able to feed superfluous power to the grid. Subsystems without generation or storage device are covered by setting their generation or battery capacity to zero, respectively.

At the current time step~$k$, the initial conditions are given by 
\begin{equation}\label{eq::initial}
	x_i(k) = \hat{x}_i\;,\;i\in[1:\mathcal{I}],
\end{equation}
with $\hat{x}_i \in [0,C_i]$. Note that the future net consumption $w_i(n)$, $n \geq k$, is not known in advance. However, we assume that it can be predicted over the subsequent $N \in \mathbb{N}_{\geq2}$ time steps and call~$N$ the prediction horizon. Then, we define the feasible sets
\begin{equation}
\mathbb{U}_i:=\left\{u_i \in \mathbb{R}^{2N} \left| 
\begin{matrix}
\exists\;x_i(k)\;, \ldots ,\; x_i(k+N-1) \in \mathbb{R}:\\
\text{initial condition~\eqref{eq::initial} holds,}\\
\text{system dynamics~\eqref{eq::dynRES1} and}\\
\text{constraints~\eqref{subeq:constraints} hold for all}\\
\text{time instants } n \in [k:k+N-1]\end{matrix}
\right\}\right. \nonumber
\end{equation}
for all $i \in [1:\mathcal{I}]$ with stacked control inputs $$u_i=(u_i(k)^\top\,\ldots u_i(k+N-1)^\top)^\top\;.$$
Note that the sets $\mathbb{U}_i$ and hence, $\mathbb{U} = \mathbb{U}_1 \times \ldots \times \mathbb{U}_\mathcal{I}$ are (convex) polytopes. 

\subsection{Optimal peak shaving}
From a grid operator's point of view it is desirable to provide a constant control energy. Therefore, our goal is to flatten the aggregated power demand profile. To this end, the deviation of the average power demand from a desired reference trajectory $\bar{\zeta} = (\bar{\zeta}(k)\,\ldots \bar{\zeta}(k+N-1))^\top \in \mathbb{R}^N$ is penalized, i.e.,
\begin{align}\notag
&\frac{1}{N}\sum_{n=k}^{k+N-1}\left\|\frac{1}{\mathcal{I}} \sum_{i=1}^{\mathcal{I}} z_i(n) - \bar{\zeta}(n)\right\|_2^2\\\notag
\overset{\eqref{eq::dynRES2}}{=}\;&\frac{1}{N}\sum_{n=k}^{k+N-1}\left\|\frac{1}{\mathcal{I}}  \sum_{i=1}^{\mathcal{I}} \left(u_i^+(n) + \gamma_i u_i^-(n)\right) +\bar{w}(n)- \bar{\zeta}(n) \right\|_2^2
\end{align}
with averages
\[
\bar{z}(n) = \frac{1}{\mathcal{I}} \sum_{i=1}^{\mathcal{I}} z_i(n)\;\text{ and }\;\bar{w}(n)=\frac{1}{\mathcal{I}}\sum_{i=1}^{\mathcal{I}}w_i(n)\;,
\]
$n \in [k:k+N-1]$. Assuming $k \geq N-1$, we choose~$\bar{\zeta}$ to be the overall average net consumption
\begin{equation}\notag
\bar{\zeta}(n) = \frac{1}{N} \sum_{j=n-N+1}^n \bar{w}(j)\;.
\end{equation}
By introducing matrices 
\begin{align}
A_i = \frac{1}{\mathcal{I}} \cdot I_N \otimes \begin{pmatrix} 1 & \gamma_i \end{pmatrix}\in \mathbb{R}^{N \times 2N} \,,\; i \in [1:\mathcal{I}] \nonumber
\end{align}
and the vector $b = \bar{\zeta} - \bar{w} \in \mathbb{R}^N$ the objective function can be written as 
\begin{align}\label{eq::obj_peak_shaping}
	\frac{1}{N}\left\|\sum_{i=1}^{\mathcal{I}}A_iu_i - b\right\|_2^2\;.
\end{align}
In the context of group sparsity each $u_i \in \mathbb{R}^{2N}$ can be considered a group of $u=(u_1^\top\,\ldots\,u_\mathcal{I}^\top)^\top \in \mathbb{R}^{2N\mathcal{I}}$. In the following, we will enforce group-sparsity of~$u$ by introducing an $\ell_{p,1}$ regularization. 

\subsection{Group sparse control of batteries}
\begin{figure}[htbp!]
\centering
\subfigure[$\ell_{2,1}$]{ 
	\label{fig:subfig:a} 
	\includegraphics[width=0.22\textwidth]{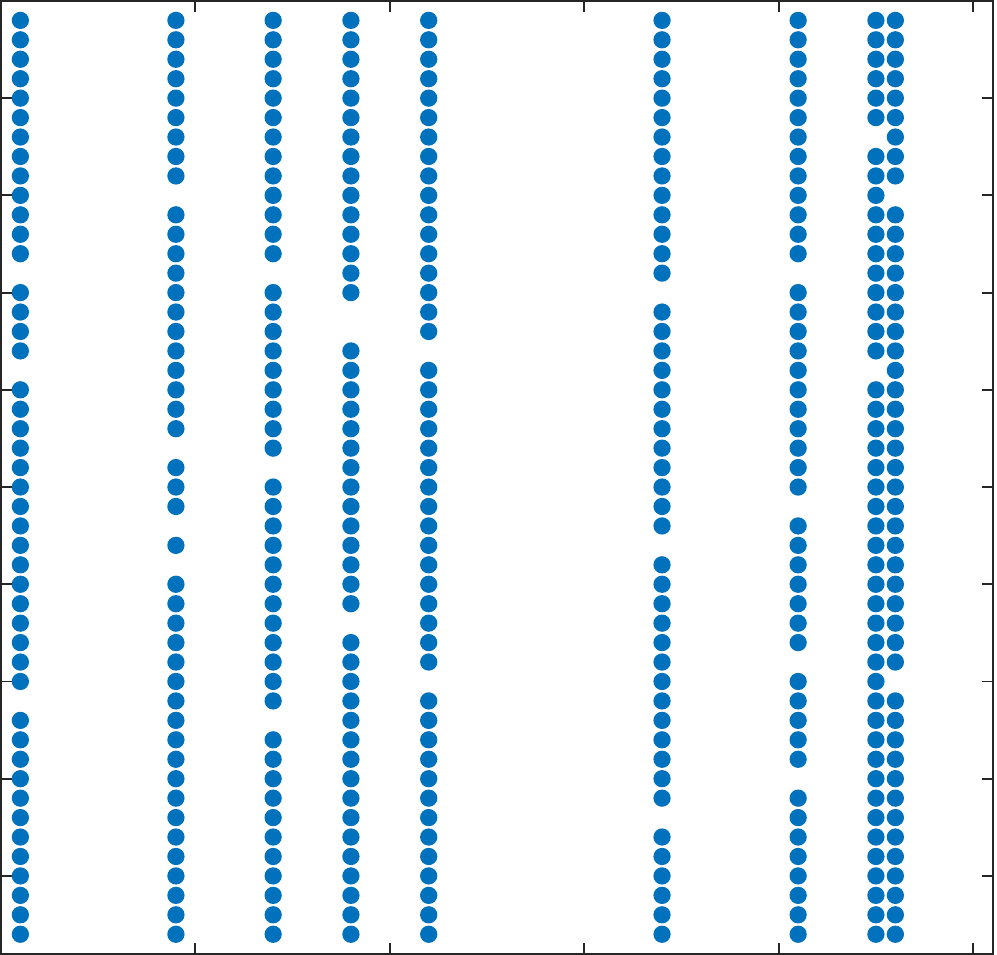}}
\hspace{0.2cm} 
\subfigure[$\ell_{1,1}$]{ 
	\label{fig:subfig:b} 
\includegraphics[width=0.22
\textwidth]{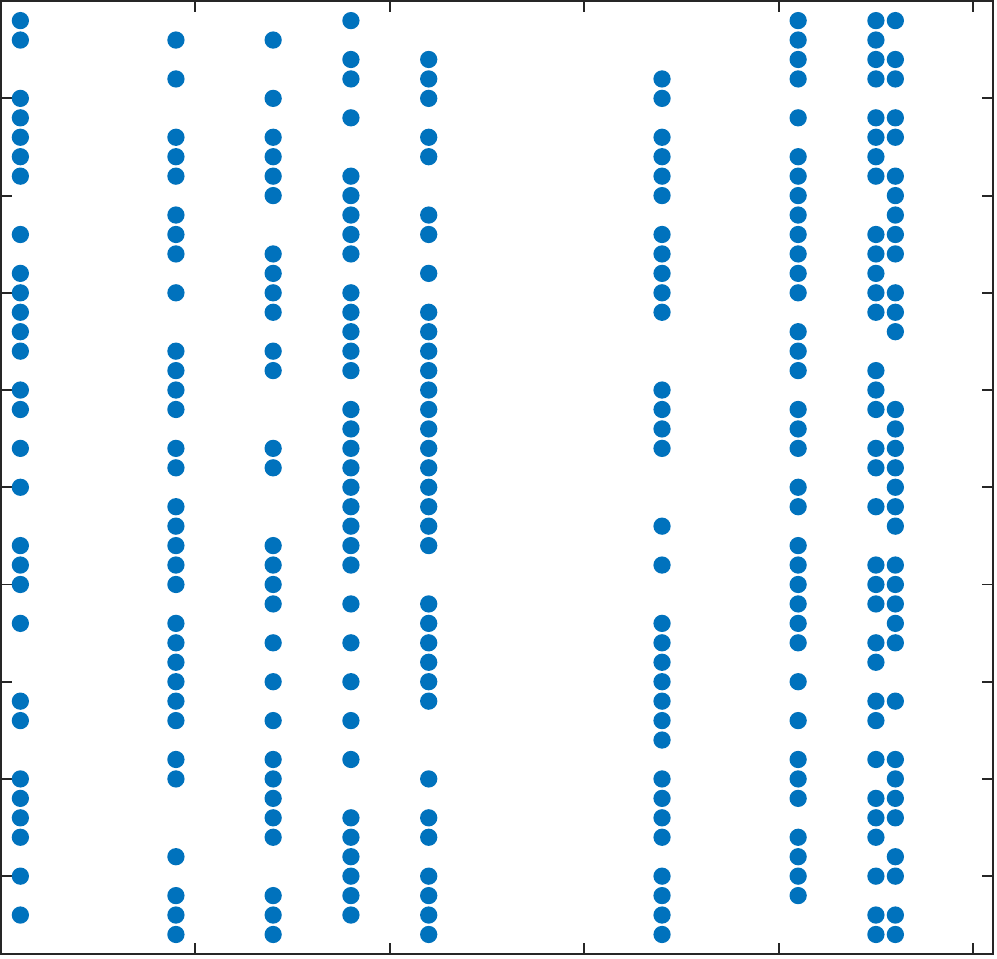}}
\caption{Group sparsity patterns based on $\ell_{p,1}$ with $p=2,1$.}
\label{fig::sparse_patterns}
\end{figure}

Optimizing~\eqref{eq::obj_peak_shaping} with respect to all feasible $u \in \mathbb{U}$ typically results in perpetual charging and discharging of the batteries~(\cite{Braun2018}). We counteract this phenomenon by establishing group-sparse control, i.e., only a few batteries are active at each time step. To this end we use the weighted mixed $\ell_{p,1}$ norm, i.e., 
\[
\begin{split}
\|u\|_{p,1}^* 
& \;=\; \sum_{i=1}^\mathcal{I} \sigma_i \|u_i\|_p 
\\[0.12cm]
& \; = \; \sum_{i=1}^\mathcal{I} \sigma_i \left(\sum_{n=k}^{k+N-1} \left(\left|u_i^+(n)\right|^p + \left|u_i^-(n)\right|^p\right) \right)^{1/p}
\end{split}
\]
with non-negative weights $\sigma = (\sigma_1\,\ldots\,\sigma_\mathcal{I})^\top \in \mathbb{R}^\mathcal{I}$ and $1\leq p <\infty$~(\cite{Yuan2006,Hu2017}). In this paper, we consider $p \in \{1,2\}$, see Fig.~\ref{fig::sparse_patterns} for a visualization of the corresponding sparsity patterns. Here, each column represents a group, i.e., in our case the control $u_i$ of the $i$-th battery. It can be seen that both norms induce group sparsity. Compared to the $\ell_{2,1}$ case, the $\ell_{1,1}$ norm additionally enforces each non-zero group to be sparse. Hence, the total number of non-zero components is further reduced. 

The main idea of exploiting group sparsity in smart grids is that at a fix time instant $n \in \mathbb{N}_0$ only a few batteries are used to manipulate the power demands.
In practice, if the net consumption is below the reference trajectory, i.e., $\bar{w}(n) < \bar{\zeta}(n)$, we create an artificial demand by charging some batteries to compensate the gap, e.g. $u_i^+(n) > 0$ for some $i$ and $u_i^-(n) = 0$ for all $i$, $i \in [1:\mathcal{I}]$. Enforcing sparse control using $\ell_{2,1}$ or $\ell_{1,1}$ regularization yields that only the batteries with the most efficient charging rates are used. Analogous argumentation holds true for $\bar{w}(n) > \bar{\zeta}(n)$. 
To avoid this one-sided usage of the batteries, the weights $\sigma$ are required to be updated online. 
In the context of MPC, this yields that the problem setup and thus the resulting group sparsity pattern vary. 

\subsection{Problem formulation}
Based on the considerations of the previous subsections we can have the group sparse control problem 
\begin{equation}\label{eq::mpc}
	\begin{split}
	\min_{u}&\;\;\frac{1}{N}\left\|\bar{z}-\bar{\zeta}\right\|_2^2 +\kappa\sum_{i=1}^{\mathcal{I}}\left(\sigma_i \|u_i\|_p\right)\\
	\text{s.t.}&\;\;\bar{z} = \sum_{i=1}^{\mathcal{I}} A_iu_i+ \bar{w}\\
	&\;\;u_i\in\mathbb{U}_i\;,\;i\in [1,\mathcal{I}]\;,
	\end{split}
\end{equation}
where, the parameter $\kappa>0$ denotes the trade-off between the optimal peak shaving and joint sparse activity of the batteries.
Here, the stage cost of each battery given by $\ell_p$ norm makes $u_i(n)$, $n\in[k,k+N-1]$ be coupled  when $p>1$. This differs from the classical MPC formulation, which takes additive stage costs with respect to time steps into account.  
By substituting~\eqref{eq::obj_peak_shaping} into~\eqref{eq::mpc}, we rewrite~\eqref{eq::mpc} into a standard composite optimization form
\begin{equation}\label{eq::probSparseRef}
	\begin{split}
	\min_{v,u} &\;\;\frac{1}{N}\left\| \sum_{i=1}^{\mathcal{I}}A_iv_i-b \right\|_2^2 +  \sum_{i=1}^{\mathcal{I}}\tilde{\sigma}_i\left\| u_i \right\|_{p} \\[0.12cm]
	\text{s.t.} &\;\;\left\{
	\begin{array}{l}
	v_i=u_i\;\; \mid\lambda_i\;,\;i\in[1:\mathcal{I}]\\[0.16cm]
	u_i\in\mathbb{U}_i\;,\;i\in[1:\mathcal{I}]
	\end{array}
	\right.
	\end{split}
\end{equation}
with $\tilde{\sigma}_i=\kappa\cdot\sigma_i$.
Here, we introduce an auxiliary variable $v\in\mathbb{R}^{2N \mathcal{I}}$ and denote by $\lambda_i$ the Lagrangian multipliers of the constraints $v_i=u_i$. Note that, operator splitting methods have already been developed for solving~\eqref{eq::probSparseRef} in a distributed manner~(\cite{Boyd2011}). In the next section, we will propose a distributed model predictive control scheme to solve~\eqref{eq::probSparseRef} by using the state-of-the-art ADMM method. 

\section{Distributed Sparse Optimization}\label{sec:distributed_MPC}

In this section we elaborate how to solve~\eqref{eq::probSparseRef} in a distributed MPC scheme. First, Algorithm~\ref{alg::admm} outlines how to solve~\eqref{eq::probSparseRef} via ADMM in a distributed manner. The subsystems solve decoupled small-scale problems in parallel and the CE solves an unconstrained Quadratic Programming~(QP). Then, the proposed method is embedded into a model predictive control scheme reflected in Algorithm~\ref{alg::mpc}.
 
\subsection{Alternating Direction Method of Multipliers}
Referring to~(\cite{Boyd2011}), Algorithm~\ref{alg::admm} with four main steps outlines to solve~\eqref{eq::probSparseRef} by using ADMM.
\begin{algorithm}[htbp!]
	\caption{ADMM for solving~\eqref{eq::probSparseRef}}
	\textbf{Input:} initial guesses $(u^0,v^0,\lambda^0)$ and step size $\rho^0>0$, stop tolerance $\varepsilon>0$, tuning parameter $\eta,\mu>0$. \\[0.16cm]
	\textbf{For }$m=0:\mathrm{MaxIte}$
	\vspace{0.1cm}
	\begin{itemize}
		\item[1)] \textit{Parallel Step:} Compute for all $i\in[1:\mathcal{I}]$ in parallel
		\[
		\begin{split}
		u_i^{m+1} \;=\;&\argmin_{u_i\in\mathbb{U}_i}\;\; \tilde{\sigma}_i\|u_i\|_{p} + \frac{\rho^m}{2}\left\|u_i-\frac{\lambda_i^m}{\rho^m}-v_i^m\right\|_2^2\\[0.12cm]
		\lambda_{i}^{m+1}\; =\;& \lambda_i^m +\rho^m(v_i^{m}-u_i^{m+1}) 
		\end{split}
		\]
		\item[2)] \textit{Consensus Step:} Solve unconstrained QP
	\end{itemize}
	\[
	\begin{split}
	\;v^{m+1} =&\,\argmin_v\frac{1}{N}\|Av-b\|_{2}^2 + \frac{\rho}{2}\left\|v-u^{m+1}+\frac{\lambda^{m+1}}{\rho}\right\|_2^2\,\\
	=&\,\left( \frac{2}{N}A^\top A + \rho^m I\right)^{-1}\\ &\qquad \qquad \cdot \left(\frac{2}{N}A^\top b - \lambda^{m+1} + \rho^m u^{m+1}\right)	
	\end{split}		
	\]
	\begin{itemize}
		\item[3)] \textit{Stop Criterion:} Evaluate 
		\[
		r^\mathrm{pri} = \rho\|u^{m+1}-v^{m+1}\|_2\;,\;r^\mathrm{dual}=\rho\|(v^{m+1}-v^m)\|_2.
		\]
		If $r^\mathrm{pri}\leq \varepsilon$ and $r^\mathrm{dual}\leq \varepsilon$, then terminate. 
		\item[4)] \textit{Adpative Dual Step Size:} Update $\rho^{m+1}$ by 
		\[
		\rho^{m+1}\leftarrow \left\{
		\begin{array}{cl}
		\eta \rho^{m}&\text{if }r^\mathrm{pri}\geq \mu r^\mathrm{dual}\\[0.12cm]
		\rho^{m}/\eta&\text{if }r^\mathrm{dual}\geq \mu r^\mathrm{primal}\\[0.12cm]
		\rho^{m}&\text{otherwise}\
		\end{array}
		\right.
		\]
	\end{itemize}
	
	\textbf{End}
	\label{alg::admm}
\end{algorithm}
In the first step the local primal variable $u_i$ and dual variable $\lambda_i$ are updated in parallel. Then, we solve an unconstrained QP in the consensus step. Note that the solution map is worked out analytically. In order to check the terminal condition, the primal and dual residual $r^\mathrm{pri}$, $r^\mathrm{dual}$ are evaluated in Step~3. Here, in contrast to~(\cite{Boyd2004}) we do not use the relative tolerance but a fixed $\varepsilon > 0$. In order to speed up the convergence, we utilize an adaptive strategy to update the dual step size~$\rho^m$. This heuristic increases $\rho^m$ if $r^\mathrm{pri}$ decreases faster than $r^\mathrm{dual}$ and vice versa, see e.g.~(\cite{Boyd2011}) for a possible choice of the tuning parameters~$\eta$ and~$\mu$. Algorithm~\ref{alg::admm} requires the grid operator to collect $4N$ floats information from each subsystem and spread $2N+1$ back to the subsystems per iteration. Note that the grid operator does not require to have any information on the local system model.

\subsection{Local solver}
At the parallel step of Algorithm~\ref{alg::admm}, the $u_i$ update requires to solve a constrained lasso problem. We propose two efficient local solvers to update~$u_i$ depending on the case $p \in \{1,2\}$. 
\begin{enumerate}
	\item If $p=1$, the $\ell_1$ term in the objective can be reformulated into the constraints~(\cite{Boyd2004}) by introducing auxiliary variables $s_i\in\mathbb{R}^{2N}$. This yields the decoupled QP
	\begin{equation}
	\begin{split}
	\min_{s_i,u_i} \quad & \mathbf{1}^\top s + \frac{\rho^m}{2}\left\|u_i-\frac{\lambda_i^m}{\rho^m}-v_i^m\right\|_2^2\\
	\text{s.t.} \quad &u_i\in\mathbb{U}_i\;,\;s \leq  u\leq  s\;,
	\end{split}
	\end{equation}
	which allows for direct usage of existent QP solvers such as \texttt{qpOASES}~(\cite{Ferreau2014}). 
	\vspace{0.2cm}
	\item If $p=2$, we propose to use a local ADMM solver as follows,
	\begin{equation}
	\begin{split}
	s_i \;=\;&\mathcal{S}_{\tilde{\sigma}_i/\rho^m}\left(
	v_i^m + u_i^j+\frac{\lambda_i^m-\xi^j}{\rho^m}
	\right)\;, \\
	u_i^{j+1}\;=\;&\argmin_{u_i\in\mathbb{U}_i}\;\;  \frac{\rho^m}{2}\left\|u_i-s_i-\frac{\xi_i^j}{\rho^m}\right\|_2^2\;, \\
	\xi_i^{j+1}\;=\;&\xi_i^j + \rho^m (u_i^{j+1}-s_i)\;,
	\end{split}
	\end{equation}
	where superscript $j$ represents the iteration of the inner ADMM loop, $\mathcal{S}_a : \mathbb{R}^{2N} \to \mathbb{R}^{2N}$ denotes the \emph{soft thresholding operator} defined by
	\[
	\mathcal{S}_a(x) = 
	\max\left\{1 - a/\left\| x \right\|_2,0\right\} x \;.
	\]
	Here, the omitted terminal condition is analogous to Step~3) in Algorithm~\ref{alg::admm} and a fixed dual step size consistent with the current $\rho^m$ is applied. 
\end{enumerate}

\subsection{Distributed predictive sparse control}
The model predictive control scheme requires to solve~\eqref{eq::probSparseRef} during each sampling time based on the current measurements. Embedding Algorithm~\ref{alg::admm}, Algorithm~\ref{alg::mpc} outlines an ADMM based distributed predictive sparse control scheme. 
\begin{algorithm}[htbp!]
	\caption{Distributed predictive control scheme}
	\textbf{Offline:}
	\begin{itemize}
		\item Initial guess $(u^0,\lambda^0)$, set $k=0$, $v^0=u^0$ and choose weights~$\sigma$ and a tolerance $\varepsilon>0$. 
	\end{itemize}
	\textbf{Online:}
	\begin{itemize}
		\item[1)]  \textbf{Subsystems} measure current SoC $x_i(k)$, predict future net consumption $w_i$ and send it to grid operator.
		\item[2)] \textbf{Grid Operator} computes the reference trajectory $\bar \zeta$.
		\item[3)] Optionally update weights $\sigma$. Run Algorithm~\ref{alg::admm} for solving~\eqref{eq::probSparseRef} to obtain $u^*$ and $\lambda^*$. 
		\item[4)] \textbf{Subsystems} apply $u_i^*(k)$ if $\|u_i(k)\|_2\geq \varepsilon$ and apply $0$ otherwise for $i\in[1:\mathcal{I}]$.
		\item[5)] Reinitialize  
		\[
		\begin{split}
		u_i^0\;=\;& (u_i^*(k+1)^\top\,\ldots\,u_i^*(k+N-1)^\top\;0_2^\top)^\top\\
		\lambda_i^0 \;=\;& (
		\lambda^*_i(k+1)\,\ldots\,\lambda^*_i(k+N-1)\;0
		)^\top
		\end{split}
		\]
		for all $i\in[1:\mathcal{I}]$.
		Then, set $k\leftarrow k+1$ and go to Step~1).
	\end{itemize}
	\label{alg::mpc}
\end{algorithm}

In order to achieve different sparsity patterns in each time step we optionally update the weights~$\sigma$ in Step~3). In practice, we run Algorithm~\ref{alg::admm} to a predetermined numerical accuracy such that we choose a tolerance for applying the sparse control at Step~4) in  Algorithm~\ref{alg::mpc}. Step~5) in the Algorithm~\ref{alg::mpc} is a warm-start step, which improves the online convergence performance of Algorithm~\ref{alg::admm}~(\cite{Braun2018}). In the following section, we will illustrate the numerical performance of Algorithm~\ref{alg::mpc} by applying it to benchmark problems.

\section{Numerical Results}\label{sec:numerics}

In this section we compare the numerical results of $\ell_{1,1}$ and $\ell_{2,1}$ norm for both open-loop and closed-loop control. To this end we consider heterogeneous systems with randomly generated parameters according to Table~\ref{tab:parameters}. 
Furthermore, we set $T = 0.5$ [h], $N = 24$, and $\hat{x}_i = 0.5$~[kWh] for all $i \in [1:\mathcal{I}]$. 
\begin{table}[htbp!]
\centering
\caption{Parameters for implementation.}
\renewcommand{\arraystretch}{1.2}	
\begin{tabular}{|c||c||c|}
\hline 
& expected value & standard deviation \\
\hline
$C_i$ & $2.0563$ [kWh] & $0.2431$ [kWh] \\
\hline 
$\overline{u}_i$ & $0.5229$ [kW] & $0.1563$ [kW] \\
\hline
$\underline{u}_i$ & $-0.5105$ [kW] & $0.1474$ [kW] \\ 
\hline 
$\alpha_i$ & $0.9913$ & $0.0053$ \\
\hline
$\beta_i$ & $0.9494$ & $0.0098$ \\
\hline
$\gamma_i$ & $0.9487$ & $0.0100$ \\
\hline
\end{tabular} 
\label{tab:parameters}
\end{table}

\subsection{Open-loop optimal control}
Let us have a closer look at the open-loop performance in this section.
Table~\ref{tab:impact_I} illustrates the impact of the number of systems~$\mathcal{I}$ for the solution sparsity with fix $\kappa = 10^{-3}$. We ran a case study for different weights~$\sigma$ and listed the mean value, the standard deviation, and the median of the percentage of the non-zero components of the optimal control. One can see that the larger the grid the higher the sparsity rate. 
\begin{table}[htbp!]
	\centering
	\caption{Impact of the number of subsystems~$\mathcal{I}$ on the percentage of non-zero components of the optimal control $u \in \mathbb{R}^{2N\mathcal{I}}$ for $\kappa = 10^{-3}$.}
	\renewcommand{\arraystretch}{1.2}	
	\begin{tabular}{|c||c|c|c|c|}
	\hline
	& $\mathcal{I}$	&	mean	&	std dev		&	median	\\
	\hline
	$\ell_{2,1}$ &  25	&	30.67	&	0.85	&	30.33	\\
	&  50	&	24.41	&	0.46	&	24.17	\\
	& 100	&	18.73	&	0.26	&	18.75	\\
	\hline
	$\ell_{1,1}$ &  25	&	8.48	&	0.36	&	8.38	\\
	&  50	&	4.65	&	0.10	&	4.65	\\
	& 100	&	3.82	&	0.06	&	3.81	\\
	\hline
	\end{tabular}
	\label{tab:impact_I}
\end{table}

\begin{figure}[htbp!]
\centering
	\includegraphics[width=0.42\textwidth]{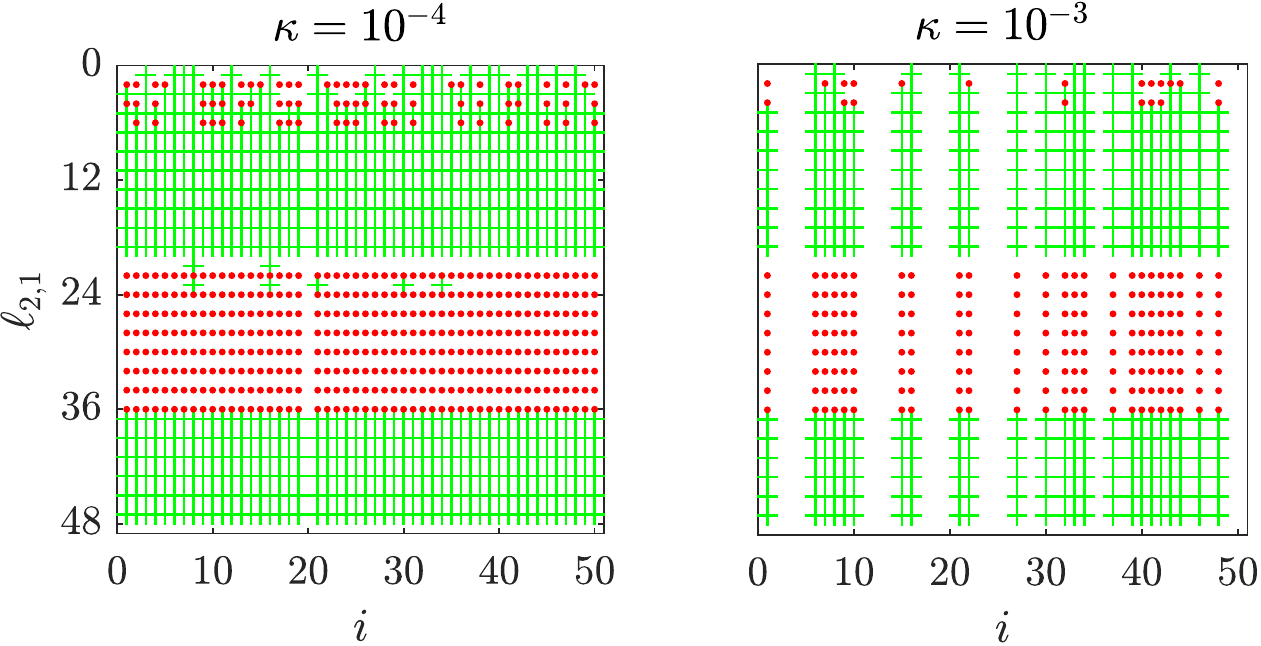}
	\\
	\includegraphics[width=0.42\textwidth]{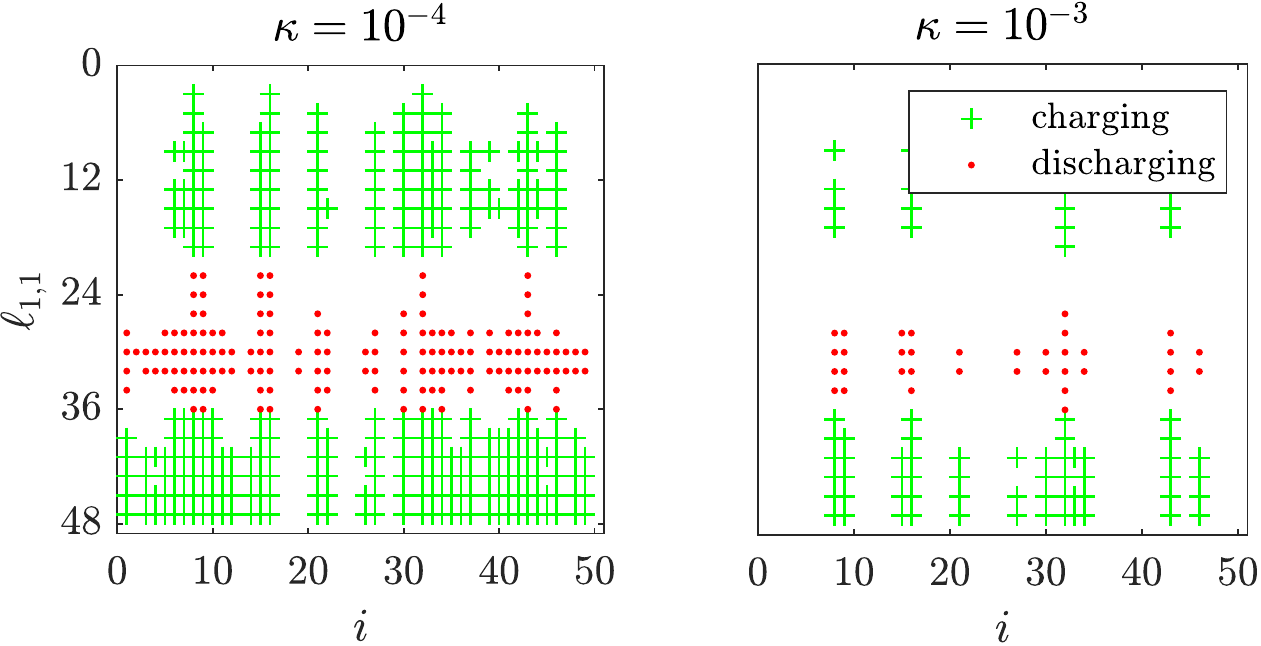}
\caption{Impact of the choice of~$\kappa$ and the regularization on the open-loop sparsity pattern for $\mathcal{I} = 50$ systems. The y-axis denotes the components of the control vector $u_i \in \mathbb{R}^{2N}$.}
	\label{fig:OL_sparsity_pattern}
\end{figure}

In the following, we fix $\mathcal{I} = 50$. Therefore, the total number of control variables is~$2400$. 
Figure~\ref{fig:OL_sparsity_pattern} visualizes the impact of the choice of the regularization and the size of the weighting parameter $\kappa \in \{10^{-4},10^{-3}\}$ on the open-loop sparsity pattern. 
Increasing~$\kappa$ the batteries are used less often. For sufficiently large~$\kappa$ some stay even inactive over the whole prediction horizon. Note that the $\ell_{1,1}$ regularization enforces $u_i^+(n) \cdot u_i^-(n) = 0$ for all $i \in [1:\mathcal{I}]$, $n \in [k:k+N-1]$, i.e., only charging or discharging at one time instant. This is not the case if an $\ell_{2,1}$ regularization is used. 
\begin{figure}[htbp!]
	\centering
	\subfigure[$\ell_{2,1}$]{ 
		\label{fig:subfig:OL_dev_L2} 
		\includegraphics[width=0.23\textwidth]{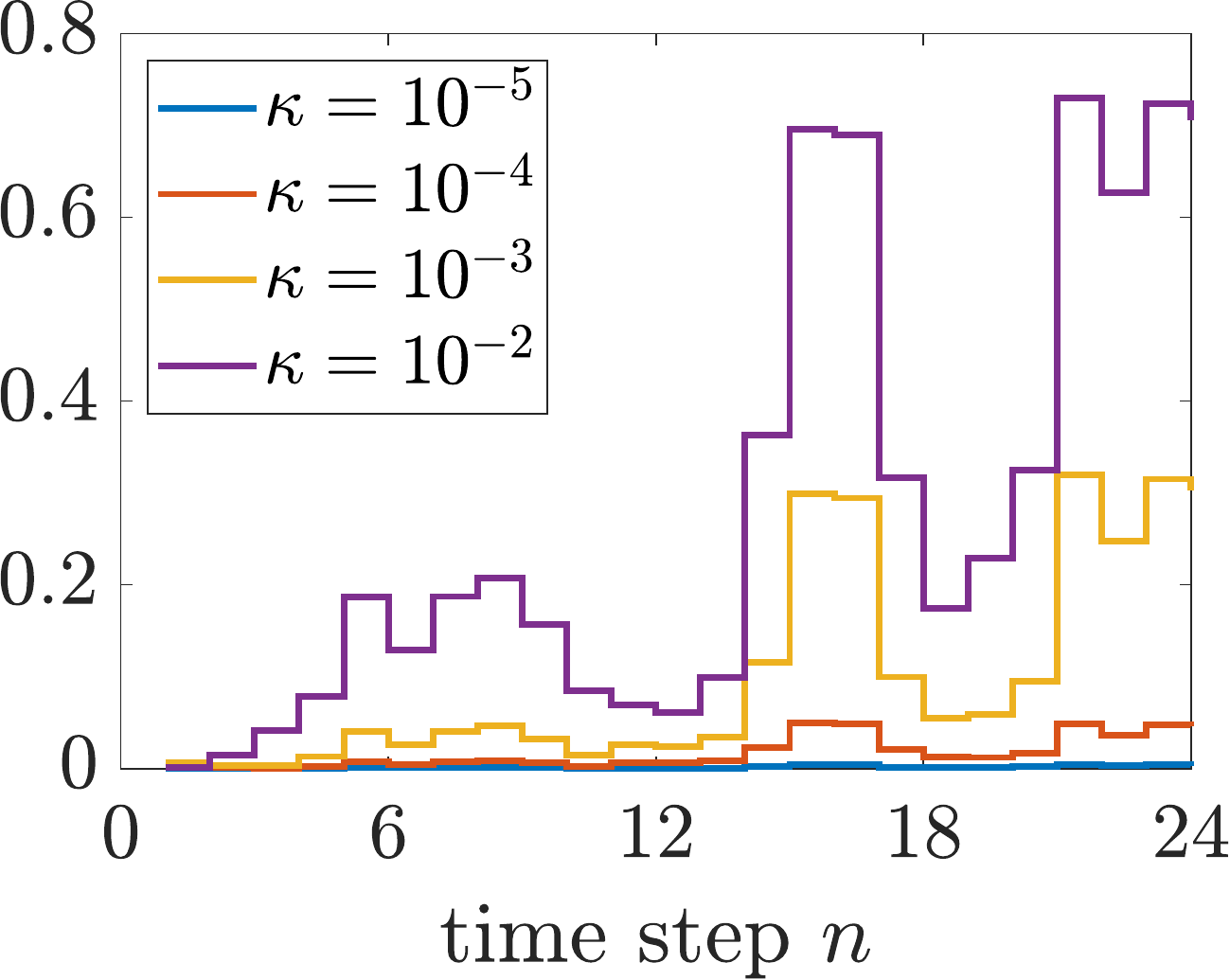}}
	\subfigure[$\ell_{1,1}$]{ 
		\label{fig:subfig:OL_dev_L1} 
		\includegraphics[width=0.23\textwidth]{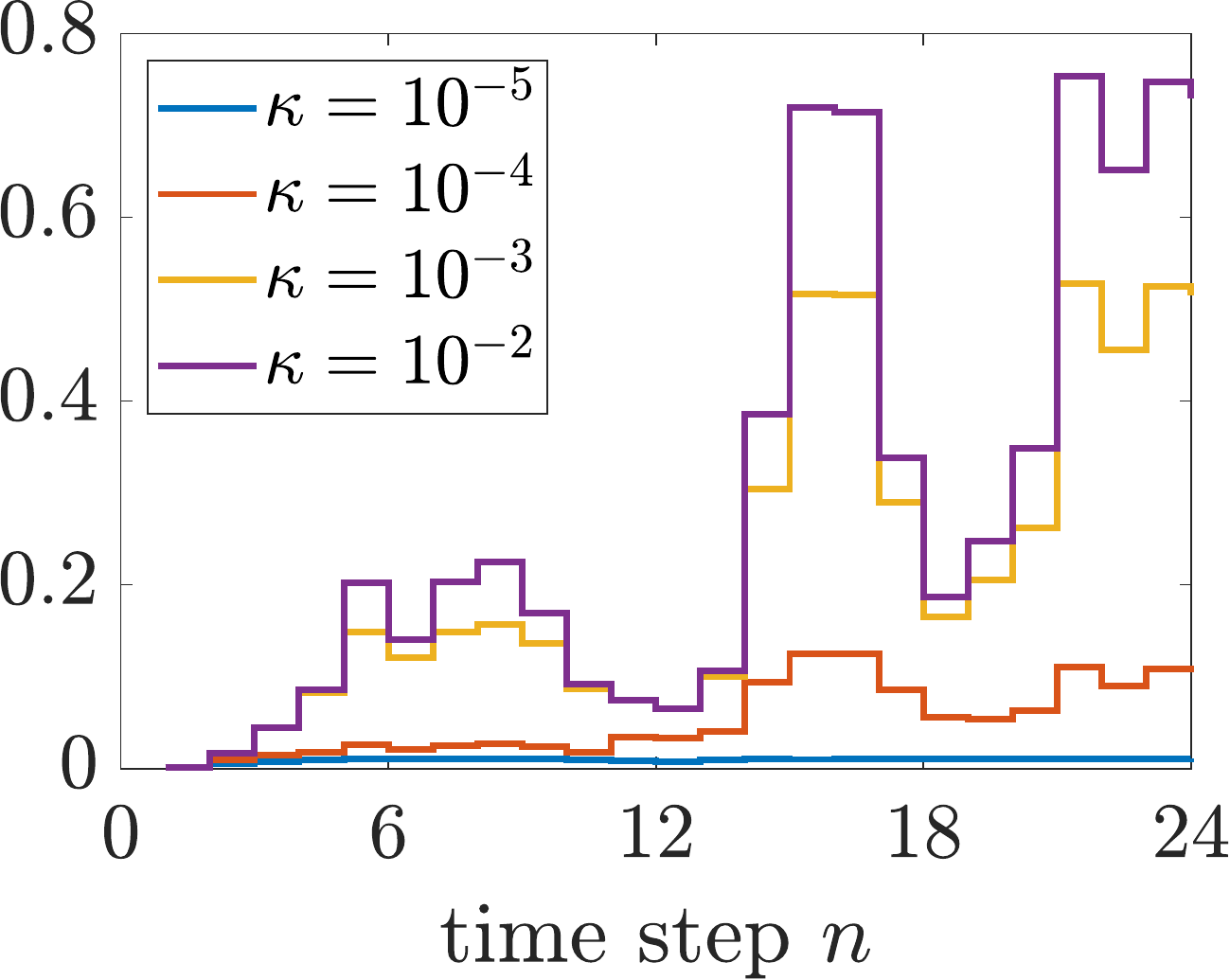}}
	\caption{Relative deviation $\left|\bar{z}(n;\kappa) - \bar{z}(n;0)\right| / \left\|\bar{z}(\cdot \,;0)\right\|_\infty$ of $\bar{z}(\cdot\,;\kappa)$ from $\bar{z}(\cdot\,;0)$ for $\mathcal{I} = 50$ systems.}
	\label{fig:OL_dev_from_kappa0}
\end{figure}

Figure~\ref{fig:OL_dev_from_kappa0} shows the overall performance with respect to~\eqref{eq::obj_peak_shaping} depending on $\kappa \in [10^{-5},10^{-2}]$. More precisely, the relative deviation from the solution associated with $\kappa = 0$ is depicted. The larger~$\kappa$, i.e., the less batteries are active, the worse the performance. 
The $\ell_{1,1}$ regularization with $\kappa = 10^{-4}$ achieves reasonable performance with respect to peak shaving while establishing sparsity.

\subsection{MPC closed loop}
This section illustrates the closed-loop performance of Algorithm~\ref{alg::mpc} depending on the choice of the penalty term and the weighting parameter. Note that if the weights $\sigma = \sigma(k)$ are changed in every single MPC step~$k$, group sparsity cannot be established, since in each step different devices might be active compared to the previous one. Therefore, in our implementation, we generate varying weights every three hours by using the \texttt{MATLAB} command \texttt{randn}, which yields normally distributed random numbers, i.e.,
\begin{align}
	\sigma_i(k) \sim \mathcal{N}(0,1). \nonumber
\end{align}
\begin{figure}[htbp!]
\centering
	\includegraphics[width=0.42\textwidth]{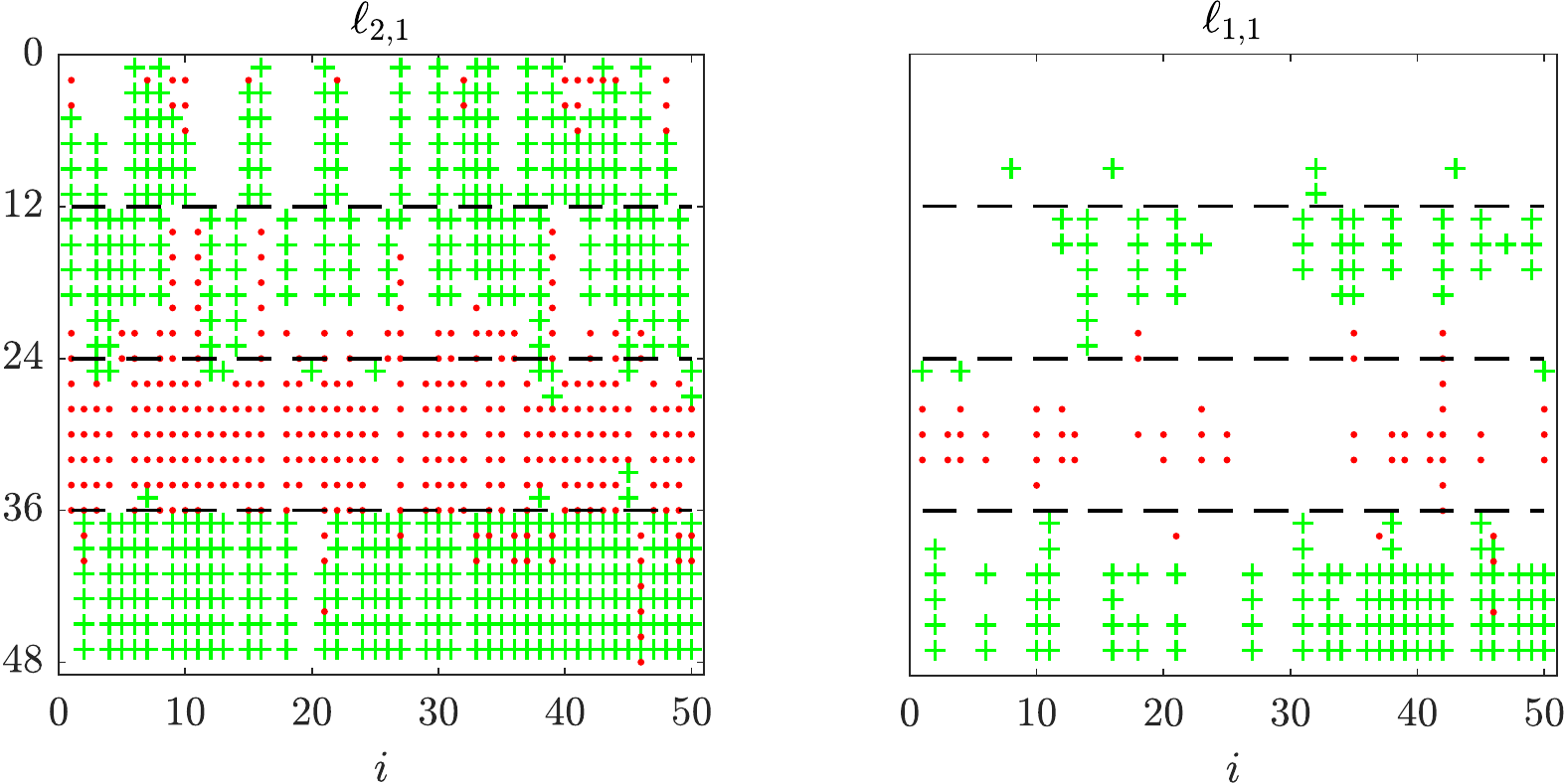}
\caption{Closed-loop sparsity pattern for $\mathcal{I} = 50$ systems and $\kappa = 10^{-3}$. The dashed black lines indicate new weights~$\sigma$.}
\label{fig:CL_sparsity_pattern}
\end{figure}
For the same reason we omit disturbances in the forecasting variables~$w_i$. The percentage of non-zero components of the optimal solution is approximately 34\% and 8\% for the $\ell_{2,1}$ and the $\ell_{1,1}$ case, respectively. Hence, similar to the open-loop simulation, the $\ell_{1,1}$ solution is sparser than the $\ell_{2,1}$ solution. 
Note that the open-loop solutions involve several devices to be inactive while others are active the whole time, see Figure~\ref{fig:OL_sparsity_pattern}. Thanks to the updated weights~$\sigma$ this phenomenon does not occur in the closed loop, see Figure~\ref{fig:CL_sparsity_pattern}. After each three hours time interval the sparsity might change.

\section{Conclusions}\label{sec:conclusions}

In this paper we considered a smart grid optimization problem dealing with optimal control of distributed energy storage devices. We proposed a distributed optimization scheme yielding group sparse optimal control. Our numerical results show that this approach is able to reduce the usage of each single battery in order to prolong their life time in a receding horizon fashion.

\bibliography{root}

\end{document}